# A Common Gene Expression Signature Analysis Method for Multiple Types of Cancer

Yingcheng Sun, Xiangru Liang and Kenneth Loparo

Case Western Reserve University, Cleveland OH 44106, USA
{yxs489, xxl487, kal4}@case.edu

**Abstract.** Mining gene expression profiles has proven valuable for identifying signatures serving as surrogates of cancer phenotypes. However, the similarities of such signatures across different cancer types have not been strong enough to conclude that they represent a universal biological mechanism shared among multiple cancer types. Here we describe a network-based approach that explores gene-to-gene connections in multiple cancer datasets while maximizing the overall association of the subnetwork with clinical outcomes. With the dataset of The Cancer Genome Atlas (TCGA), we studied the characteristics of common gene expression of three types of cancers: Rectum adenocarcinoma (READ), Breast invasive carcinoma (BRCA) and Colon adenocarcinoma (COAD). By analyzing several pairs of highly correlated genes after filtering and clustering work, we found that the co-expressed genes across multiple types of cancers point to particular biological mechanisms related to cancer cell progression, suggesting that they represent important attributes of cancer in need of being elucidated for potential applications in diagnostic, prognostic and therapeutic products applicable to multiple cancer types.

**Keywords:** Gene Expression Signature, Gene Network, Cancer.

## 1    Introduction

Cancer is known to be not just one disease, but many diseases, as evidenced by the diversity of its pathological manifestations. With the goal of clinically stratifying samples into risk groups, several gene expression biomarkers have been proposed for a large variety of cancer types [1, 2]. Most biomarkers have been identified and designed for a specific type of cancer, but it has been appreciated that there exist some unifying capabilities or "hallmarks" that can be used as tumor markers, characterizing all cancers because they exhibit similar biomolecular phenotypes [3]. Furthermore, it has been recognized that gene expression signatures resulting from analysis of cancer datasets can serve as surrogates of cancer phenotypes [4]. Therefore, it is reasonable to hypothesize that computational analysis of rich biomolecular cancer datasets may reveal signatures that are shared across many cancer types and are associated with specific cancer phenotypes, and are related to sustaining proliferative, insensitivity to anti-growth signals, metastasis, and etc [3].

Gene expression signatures that can be applied for a broad range of cancers could be highly useful in research and clinical settings. In clinics, such signatures may serve as a standard assessment for facilitating the interpretation and broad application of laboratory test results, simplifying laboratory protocols, and reducing costs. In research, these signatures may help to elucidate broadly observed biological mechanisms and possible drug targets [8].

With the availability of rich data sets from many different cancer types provides an opportunity for thorough computational data mining in search of such common patterns. Distinct algorithms and strategies have been used to identify common gene expression signatures: regulatory network [5], clustering approaches [6] and other related techniques. In this paper, we use gene co-expression network and develop a statistical method of common gene expression signature analysis for multiple types of cancer.

Previous research evaluates such common patterns within a couple of cancer types [7, 8], but never explore the following three cancer types together: Colon adenocarcinoma (COAD), Rectum adenocarcinoma (READ) and Breast invasive carcinoma (BRCA). In this paper, we use the above three types of cancer for experiment. Our purpose is to identify gene expression pattern across multi-cancers, and reveal hallmarks of cancer, and thus helps to find bio-markers associated with different types of cancer, and contributes to the prognosis of cancer.

## 2 Method

A co-expression network identifies which genes have a tendency to show a coordinated expression pattern across a group of samples. This co-expression network can be represented as a gene–gene similarity matrix, which can be used in downstream analyses [9]. Canonical co-expression network construction and analyses can be described with the following three steps.

To begin with, one needs to define a measure of similarity between the gene expression profiles. This similarity measures the level of concordance between gene expression profiles across the experiments. Individual relationships between genes are defined based on correlation measures [29] or mutual information [10] between each pair of genes. Different measures of correlation have been used to construct networks, including Pearson's or Spearman's correlations [11]. Alternatively, least absolute error regression [12] or a Bayesian approach [13] can be used to construct a co-expression network [14]. In this paper, we use the commonly used Pearson's correlation as follows:

$$\tau = \frac{1}{n}\sum_{i=1}^{n}(\frac{x_i - \bar{x}}{\sigma_x})(\frac{y_i - \bar{y}}{\sigma_y})$$

where $\sigma_x$ is the standard deviation of x and $\sigma_y$ is the standard deviation of y.

In the second step, co-expression associations are used to construct a network where each node represents a gene and each edge represents the presence and the strength of the co-expression relationship [15]. In order to filter out the genes that are not related to cancer, we select genes expressed differently between cancer samples and normal samples with T-test. To get the common gene expression signatures, we choose the genes existed in all the types of cancer samples.

In the third step, modules (groups of co-expressed genes) are identified using one of several available clustering techniques. Clustering in co-expression analyses is used to group genes with similar expression patterns across multiple samples to produce groups of co-expressed genes rather than only pairs. The clustering method needs to be chosen with consideration because it can greatly influence the outcome and meaning of the analysis. Many clustering methods are available, such as k-means, self-organizing maps (SOM) and etc. In this paper, we use k-means as our clustering algorithm because it is faster and produce tighter clusters than other clustering methods [23, 24, 25]. Modules can subsequently be interpreted by functional enrichment analysis, a method to identify and rank overrepresented functional categories in a list of genes [16, 17].

## 3    Experiment

The datasets we used in our experiments are the level 3 data of three types of cancer from The Cancer Genome Atlas (TCGA): Rectum adenocarcinoma (READ), Breast invasive carcinoma (BRCA) and Colon adenocarcinoma (COAD). For each type of cancer, cancer samples and normal samples are randomly picked up from the same batch. Table 1 lists the number of selected genes and samples.

**Table 1.** Number of genes and samples selected

| No | Cancer | Gene Number | Caner Sample | Normal Sample |
|----|--------|-------------|--------------|---------------|
| 1 | BCRA | 17814 | 10 | 10 |
| 2 | COAD | 17814 | 10 | 10 |
| 3 | READ | 17814 | 10 | 3 |
| Total | | 53442 | 30 | 23 |

The database does not supply enough number of normal samples for READ, so we only use three normal samples as the comparison with cancer samples. Besides, all the gene expression data are detected by Agilent microarrays with the same processing and normalization way by the following formula shows, so they are comparable.

$$log_2\left(\frac{Cy5}{Cy3}\right) = log_2\left(\frac{sample}{control}\right)$$

### 3.1 Find Genes Expressed Differently between Cancer and Normal Samples

Firstly, we want to find the genes expressed differently between cancer and normal samples since these genes are more possible to be the target genes than those with the similar expressions between cancer samples and normal samples.

T-test can be used here to determine if two sets of data are significantly different from each other. Before doing T-test, we need to use F-test to examine the homogeneity of variance between two data sets. We select the data with p value larger than 0.05 after T-test. Table 2 shows the number of genes left.

**Table 2.** Number of genes left after t-test

| No | Gene | Original | After F-test | After T-test |
|----|------|----------|--------------|--------------|
| 1 | BCRA | 17814 | 12208 | 4535 |
| 2 | COAD | 17814 | 12721 | 6635 |
| 3 | READ | 17814 | 16349 | 4939 |
| Total | | 53442 | 41278 | **16109** |

### 3.2 Select Common Genes among Different Cancers

Next, we select the genes appeared in all the three gene sets and called them "common genes", because we are interested in finding the common patterns across multiple cancer types. 878 common genes are then selected from 16109 genes obtained in last step after T-test. Table 3 lists the results.

**Table 3.** Number of genes appeared in different cancer sets

| Genes | Number |
|-------|--------|
| Appear in Only One Cancers | 6676 |
| Appear in Only Two Cancers | 3398 |
| Appear in All Three Cancers | **878** |

We then list the cancer samples expressed by each gene of the three types of cancers, and obtain a gene-expression matrix with 878 rows and 30 columns, where the gene expression data of each type of cancer are represented by 10 columns of values.

### 3.3 Cluster Common Genes

With the common gene expression matrix, we can find the closely related genes with similar gene expression through clustering, which might give us clues of seeking bio-events for cancers. We use Pearson Correlation as the gene similarity metric and k-means as the clustering algorithm with the number of clusters set to 20. We use the SPSS Statistics software package (version 22) to process the data and export the analysis results. After clustering, we have 814 genes valid (92.7 %) and 64 genes missing (7.3%). It shows that most of the genes have high correlation and can be clustered into multiple groups. Fig. 1 shows the gene network.

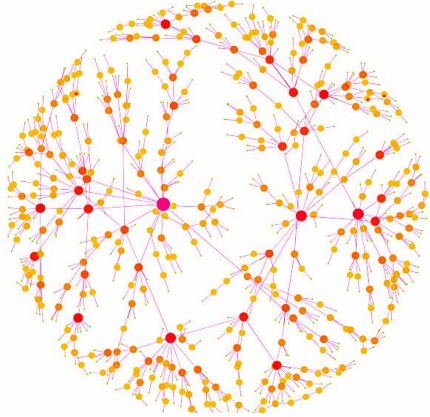

**Fig. 1.** Common gene network of selected 878 genes

In each cluster, we chose the pair of genes with highest coefficient as Table 4 shows. These genes may be related to the cancer specific bio-events, and we will discuss the significance of them in next section.

**Table 4.** Pair of genes with highest coefficient in each cluster

| Cluster | Pair of Genes | | Coefficient |
| --- | --- | --- | --- |
| Cluster1 | MT1H | MT1B | 0.984 |
| Cluster2 | PTTG3 | PTTG1 | 0.982 |
| Cluster3 | PDGFD | APOD | 0.941 |
| … | … | … | … |
| Cluster20 | FBLN5 | RBMS3 | 0.935 |

### 3.4 Cluster Samples

Besides clustering genes, we also try to find whether there are some interesting results by clustering samples. We first transpose gene-expression matrix, and then use the Pearson Correlation as the similarity metric and Between-Groups Linkage as the clustering algorithm. We only have 1 case missing and obtained a valid rate of 96.7 % for all the 30 samples. Figure 2 shows the dendrogram of clustering result. We can see that there are basically two groups of samples after clustering.

The significance of the clustering is evaluated by its statistical results. The results are basically classified into two classes: "between groups" and "within groups" values. We want the clusters with short "within group" linkage and long "between group" linkage. There are only one sample out of thirty with p value larger than 0.05:

READ-TCGA-AF-3400-01A-01R-0821-07 with 0.8. The p values of the all rest samples are less than 0.05, making the clustering significance to be 96.7%.

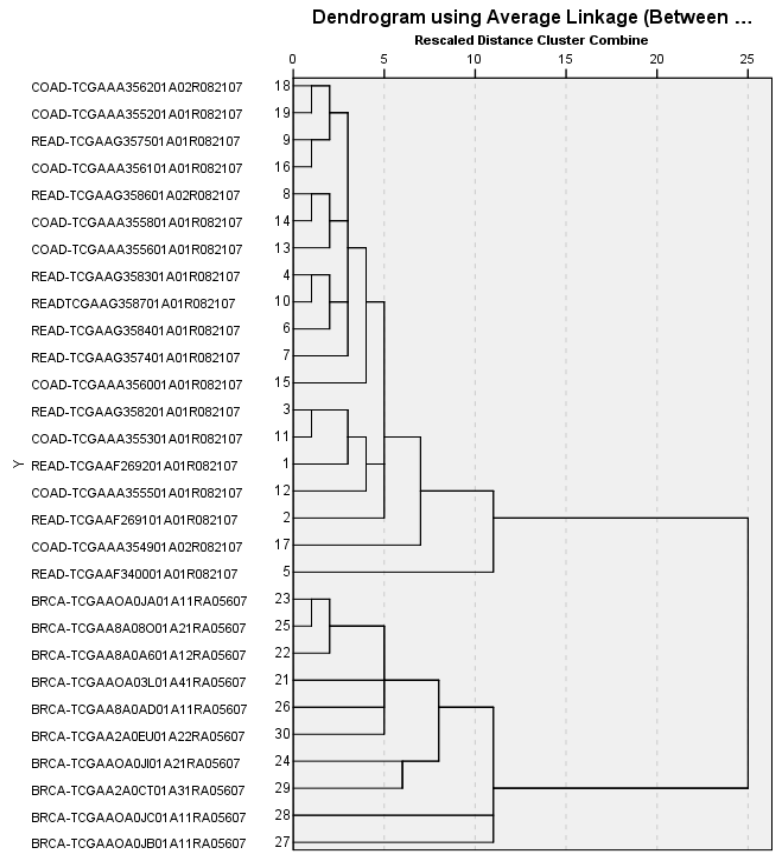

**Fig. 2.** Dendrogram of the clustering result

## 4    Analysis

We respectively choose three pairs of genes from three clusters: APOD and PDGFD, RBMS3 and FBLN5, TUBB6 and DDR2. Genes in each pair are highly related to each other, meaning their expression patterns are similar in all cases. As shown in Fig 3, each pair of genes is positively related under each type of cancer. Additionally, the values and relationships of gene expressions are closer in cancer COAD and READ, and separated from BRCA. It corresponds to the fact that rectum is a part of colon, so the origins of COAD and READ are closer, and they may share lots of commons.

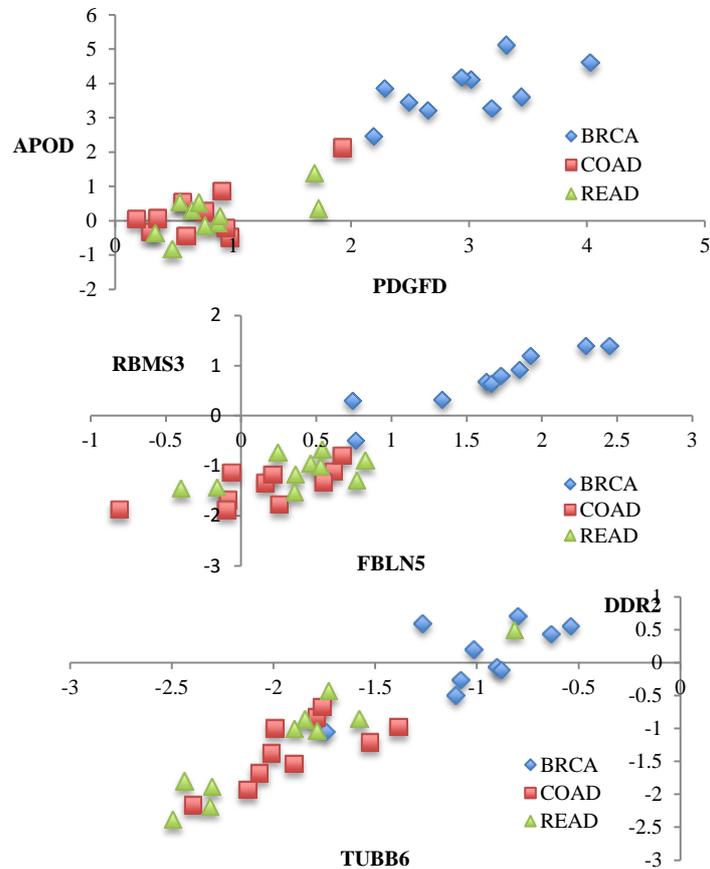

**Fig. 3.** Gene co-expression in three types of cancers

Let's focus on each pair of genes separately. In the first pair, PDGFD encodes platelet derived growth factor D, which promotes cancer progression [19]. Compared to normal samples, PDGFD is down regulated in the cancer samples. However, it makes no sense because cancer cells tend to grow rapidly and uncontrolledly, so it may up-regulate the gene expression of growth factors. APOD – apolipoprotein D is a good prognostic marker, and it is expressed differently in COAD and READ cancer samples, which may represent different patient prognosis conditions. Other genes such as FBLN5 – fibulin 5, plays a role in proliferation, migration and invasion [20], TUBB6 – tubulin beta 6, malignant transformation and drug resistance and DDR2 – discoidin domain receptor try kinase 2, which promotes matastasis. According to the other research, all these genes should contribute to cancer cells progression, but they are down regulated in the cancer samples. It may be explained by fact that patients are receiving treatments and they are turning good, so the genes promoting cancer progression are down regulated. In terms of RBMS3 – single stranded interacting protein 3 that inhibits cell proliferation is also down regulated, but it makes sense because this protein is encoded by the gene serve as a tumor suppressor. Figure 4 shows the difference of these gene expressed in normal and cancer samples.

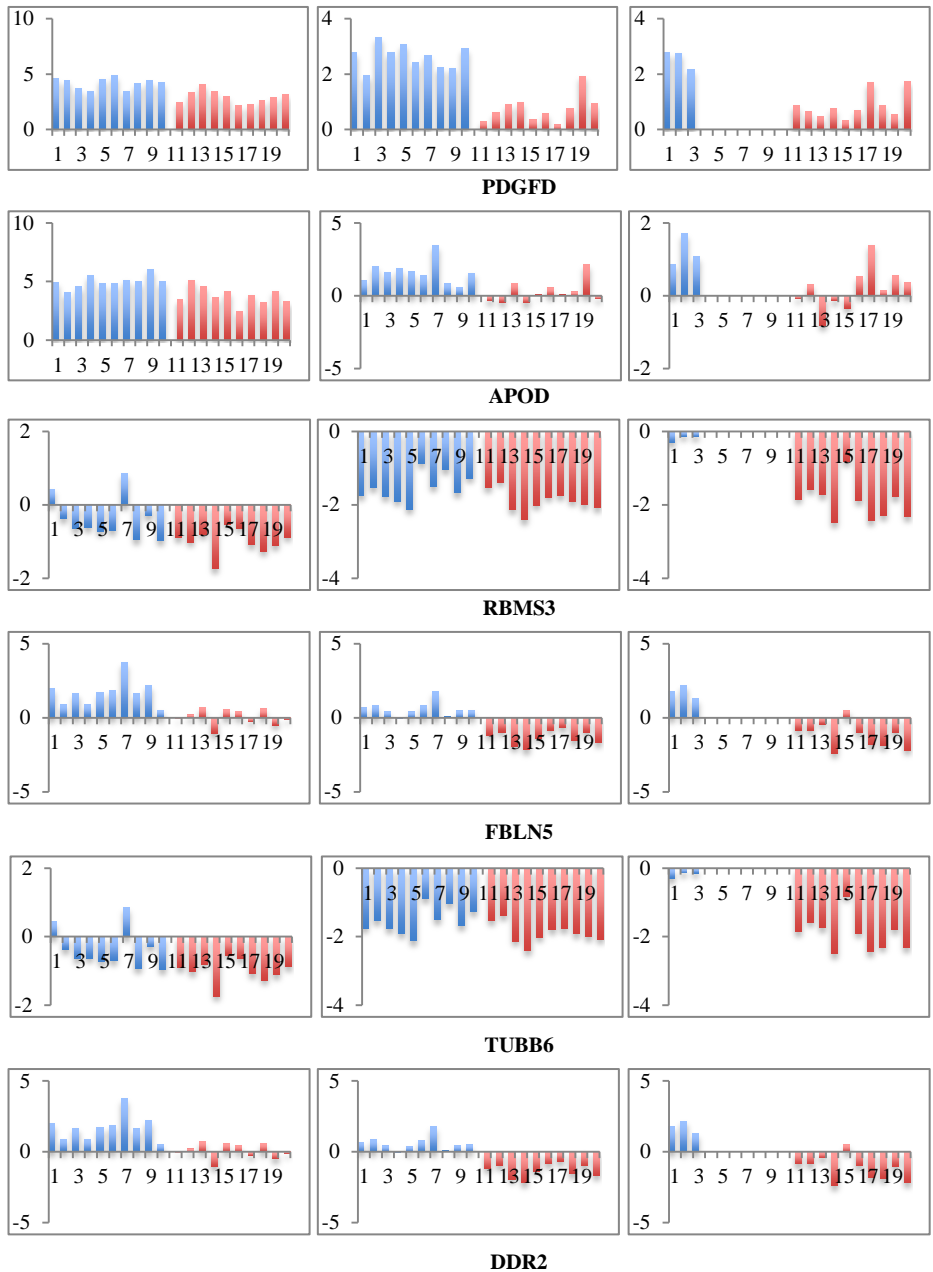

**Fig. 4.** Gene expression in normal (blue color) and cancer samples (red color) for three types of cancer: COAD, READ and BRCA

From the clustering result of sample in Figure 2, we can see that COAD and READ have closer relationships, and the patients of these two cancers mix together, while BRCA forms a separate cluster as Table 5 shows, that is an evidence of the above analysis.

**Table 5.** Clustering results of samples

| Cluster | Number of Cancer Samples | | |
|---------|------|------|------|
| | COAD | READ | BRCA |
| Cluster1 | 10 | 9 | 0 |
| Cluster2 | 0 | 0 | 10 |

## 5    Discussion

We study the real functions of the genes selected from each cluster by searching the National Center for Biotechnology Information (NCBI) database, and pay more attention to the genes that are cell essential or related to disease. In one cluster, two genes are found related to tumor suppression and migration. FLNA encodes filamin A, alpha, that is involved in remodeling the cytoskeleton to effect changes in cell shape and migration. Previous research shows that FLNA plays an important role in cancer proliferation and metastasis [18]. FLNA also interacts with oncogenesis and metastasis related proteins, meaning that it is essential for cancer progression. GAS1 is growth arrest-specific 1 which encodes GAS1 protein to blocks cells to enter S phase, so it is considered as a tumor suppress gene, and previous research suggests that GAS1 expression significantly reduce the colony-forming ability of gastric cancer cells in vitro and cell growth in vivo [21]. In Figure 5, we can see that in BRCA cancer, FLNA and GAS1 are positive related. In COAD and READ, it seems that FLNA and GAS1 do not have any relevance but their expression values are close to each other according to the dataset.

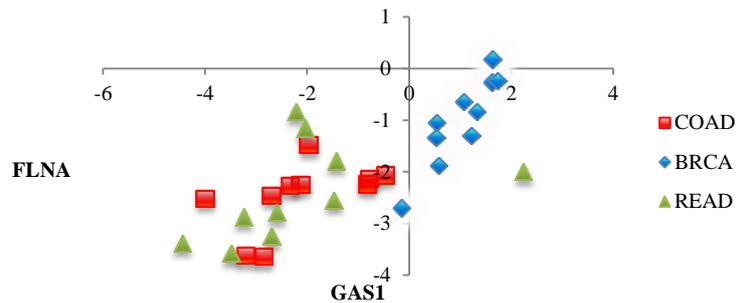

**Fig. 5.** Expression of gene FLNA and GAS1 in three types of cancers

In the future, we are interested in using machine learning methods [22, 26, 27, 31] to find more possible gene signatures, and combine the medical knowledge to explain them [28, 30].

## Acknowledgment

This work was supported by the Ohio Department of Higher Education, the Ohio Federal Research Network and the Wright State Applied Research Corporation under award WSARC-16-00530 (C4ISR: Human-Centered Big Data).

## References

1. Chen, R., Khatri, P., Mazur, P.K., Polin, M., Zheng, Y., Vaka, D., Hoang, C.D., Shrager, J., Xu, Y., Vicent, S. and Butte, A.J.: A meta-analysis of lung cancer gene expression identifies PTK7 as a survival gene in lung adenocarcinoma. Cancer research 74(10), 2892-2902 (2014)
2. Peng, Z., Skoog, L., Hellborg, H., Jonstam, G., Wingmo, I.L., Hjälm-Eriksson, M., Harmenberg, U., Cedermark, G.C., Andersson, K., Ährlund-Richter, L. and Pramana, S.: An expression signature at diagnosis to estimate prostate cancer patients' overall survival. Prostate cancer and prostatic diseases 17(1), 81 (2014).
3. Hanahan, D. and Weinberg, R.A.: Hallmarks of cancer: the next generation. cell, 144(5), 646-674 (2011)
4. Nevins, J.R. and Potti, A.: Mining gene expression profiles: expression signatures as cancer phenotypes. Nature Reviews Genetics 8(8), 601 (2007)
5. Segal, E., Friedman, N., Kaminski, N., Regev, A. and Koller, D.: From signatures to models: understanding cancer using microarrays. Nature genetics 37(6s), S38 (2005)
6. Collisson, E.A., Sadanandam, A., Olson, P., Gibb, W.J., Truitt, M., Gu, S., Cooc, J., Weinkle, J., Kim, G.E., Jakkula, L. and Feiler, H.S.: Subtypes of pancreatic ductal adenocarcinoma and their differing responses to therapy. Nature medicine 17(4), 500 (2011)
7. Cheng, W.Y., Yang, T.H.O. and Anastassiou, D.: Biomolecular events in cancer revealed by attractor metagenes. PLoS computational biology, 9(2), p.e1002503 (2013)
8. Martinez-Ledesma, E., Verhaak, R.G. and Treviño, V.: Identification of a multi-cancer gene expression biomarker for cancer clinical outcomes using a network-based algorithm. Scientific reports, 5, p.11966 (2015)
9. Van Dam, S., Vosa, U., van der Graaf, A., Franke, L. and de Magalhaes, J.P.: Gene co-expression analysis for functional classification and gene–disease predictions. Briefings in bioinformatics, 19(4), 575-592 (2017)
10. Margolin, A.A., Nemenman, I., Basso, K., Wiggins, C., Stolovitzky, G., Dalla Favera, R. and Califano, A.: ARACNE: an algorithm for the reconstruction of gene regulatory networks in a mammalian cellular context. BMC bioinformatics, Vol. 7, No. 1, p. S7 (2006)
11. Ala, U., Piro, R.M., Grassi, E., Damasco, C., Silengo, L., Oti, M., Provero, P. and Di Cunto, F.: Prediction of human disease genes by human-mouse conserved coexpression analysis. PLoS computational biology, 4(3), p.e1000043 (2008)
12. van Someren, E.P., Vaes, B.L., Steegenga, W.T., Sijbers, A.M., Dechering, K.J. and Reinders, M.J.: Least absolute regression network analysis of the murine osteoblast differentiation network. Bioinformatics, 22(4), 477-484 (2005)

13. Friedman, N., Linial, M., Nachman, I. and Pe'er, D.: Using Bayesian networks to analyze expression data. Journal of computational biology, 7(3-4), 601-620 (2000)

14. D'haeseleer, P., Liang, S. and Somogyi, R.: Genetic network inference: from co-expression clustering to reverse engineering. Bioinformatics, 16(8), 707-726 (2000)

15. Albert, R. and Barabási, A.L.: Statistical mechanics of complex networks. Reviews of modern physics, 74(1), p.47 (2002)

16. Gupta, S., Ellis, S.E., Ashar, F.N., Moes, A., Bader, J.S., Zhan, J., West, A.B. and Arking, D.E.: Transcriptome analysis reveals dysregulation of innate immune response genes and neuronal activity-dependent genes in autism. Nature communications, 5, p.5748 (2014)

17. De Magalhães, J.P., Finch, C.E. and Janssens, G.: Next-generation sequencing in aging research: emerging applications, problems, pitfalls and possible solutions. Ageing research reviews, 9(3), 315-323 (2010)

18. Zhang, K., Zhu, T., Gao, D., Zhang, Y., Zhao, Q., Liu, S., Su, T., Bernier, M. and Zhao, R.: Filamin A expression correlates with proliferation and invasive properties of human metastatic melanoma tumors: implications for survival in patients. Journal of cancer research and clinical oncology, 140(11), 1913-1926 (2014)

19. Wang, Y., Qiu, H., Hu, W., Li, S. and Yu, J.: Over-expression of platelet-derived growth factor-D promotes tumor growth and invasion in endometrial cancer. International journal of molecular sciences, 15(3), 4780-4794 (2014)

20. Hwang, C.F., Shiu, L.Y., Su, L.J., Yin, Y.F., Wang, W.S., Huang, S.C., Chiu, T.J., Huang, C.C., Zhen, Y.Y., Tsai, H.T. and Fang, F.M.: Oncogenic fibulin-5 promotes nasopharyngeal carcinoma cell metastasis through the FLJ10540/AKT pathway and correlates with poor prognosis. PloS one, 8(12), p.e84218 (2013)

21. Wang, H., Zhou, X., Zhang, Y., Zhu, H., Zhao, L., Fan, L., Wang, Y., Gang, Y., Wu, K., Liu, Z. and Fan, D.: Growth arrest-specific gene 1 is downregulated and inhibits tumor growth in gastric cancer. The FEBS journal, 279(19), 3652-3664 (2012)

22. Sun, Y., Li, Q.: The research situation and prospect analysis of meta-search engines. In: 2nd International Conference on Uncertainty Reasoning and Knowledge Engineering, pp. 224-229. IEEE, Jalarta, Indonesia (2012).

23. Li, Q., Zou, Y., Sun, Y.: Ontology based user personalization mechanism in meta search engine. In: 2nd International Conference on Uncertainty Reasoning and Knowledge Engineering, pp. 230-234. IEEE, Jalarta, Indonesia (2012).

24. Li, Q., Sun, Y., Xue, B.: Complex query recognition based on dynamic learning mechanism. Journal of Computational Information Systems 8(20), 8333-8340 (2012).

25. Li, Q., Zou, Y., Sun, Y.: User Personalization Mechanism in Agent-based Meta Search Engine. Journal of Computational Information Systems 8(20), 1-8 (2012).

26. Li, Q., Sun, Y.: An agent based intelligent meta search engine. In: International Conference on Web Information Systems and Mining, pp. 572-579. Springer, Berlin, Heidelberg (2012).

27. Sun, Y., Loparo, K.: Context Aware Image Annotation in Active Learning with Batch Mode. In: 43rd Annual Computer Software and Applications Conference (COMPSAC), vol. 1, pp. 952-953. IEEE (2019).

28. Sun, Y., Loparo, K.: A Clicked-URL Feature for Transactional Query Identification. In 2019 IEEE 43rd Annual Computer Software and Applications Conference (COMPSAC), vol. 1, pp. 950-951. IEEE (2019).

29. Sun, Y., Loparo, K.: Topic Shift Detection in Online Discussions using Structural Context. In 2019 IEEE 43rd Annual Computer Software and Applications Conference (COMPSAC) , vol. 1, pp. 948-949. IEEE (2019).

30. Sun, Y., Loparo, K.: Information Extraction from Free Text in Clinical Trials with Knowledge-Based Distant Supervision. In: IEEE 43rd Annual Computer Software and Applications Conference (COMPSAC), vol. 1, pp. 954-955. IEEE (2019).
31. Sun, Y., Loparo, K.: Learning - based Adaptation Framework for Elastic Software Systems. In: Proceedings of 31st International Conference on Software Engineering & Knowledge Engineering, pp. 281-286. IEEE, Lisbon, Portugal (2019).